\documentclass[a4paper,USenglish,cleveref, autoref, thm-restate, authorcolumns, ]{lipics-v2021}
\nolinenumbers

\usepackage{soul}

\usepackage{tikz}
\tikzstyle{mybox} = [draw=black, very thick, rectangle, rounded corners, inner ysep=5pt, inner xsep=5pt]

\usepackage{hyperref}
\usepackage{cleveref}

\usepackage{amsmath} 

\usepackage{booktabs}
\usepackage{tabularx}
\usepackage{makecell}
\usepackage{multirow}

\usepackage{tcolorbox}
\usepackage{fontawesome5}

\usepackage{caption}
\usepackage{subcaption}

\newenvironment{highlightbox}[1]{
    \begin{tcolorbox}[title={#1},
    left=2mm,right=2mm,top=1mm,bottom=1mm]
    }{
    \end{tcolorbox}
}

\crefname{figure}{Fig.}{Figs.}

\title{Human-AI Collaboration  in Requirements Engineering: Evidence of the Negative Effect of LLMs on Requirements Inspection 
} 

\titlerunning{Human-AI Collaboration in Requirements Engineering
} 

\author{Giovanna {Broccia}\footnote{corresponding author}}{CNR--ISTI, Pisa, Italy}{giovanna.broccia@isti.cnr.it}{https://orcid.org/0000-0002-4737-5761}{
}
\author{Julian {Frattini}\footnote{corresponding author}}{Chalmers University of Technology, Gothenburg, Sweden}{julian.frattini@chalmers.se}{https://orcid.org/0000-0003-3995-6125}{
}
\author{Chetan {Arora}}{Monash University, Melbourne, Australia}{}{https://orcid.org/0000-0003-1466-7386}{
}
\author{Maurice H. {ter Beek}}{CNR--ISTI, Pisa, Italy}{}{https://orcid.org/0000-0002-2930-6367}{
}
\author{Alessandro {Fantechi}}{University of Florence, Florence, Italy}{}{}{
}
\author{Andreas {Vogelsang}}{University of Duisburg-Essen}{}{https://orcid.org/0000-0003-1041-0815}{
}
\author{Alessio {Ferrari}}{Trinity College Dublin, Dublin, Ireland}{}{https://orcid.org/0000-0002-0636-5663}{
}

\authorrunning{G. Broccia et al.} 

\Copyright{Giovanna Broccia and Julian Frattini and Chetan Arora and Maurice H. ter Beek and Alessandro Fantechi and Andreas Vogelsang and Alessio Ferrari} 



\category{} 

\relatedversion{} 

\EventEditors{}
\EventLongTitle{}
\EventShortTitle{}
\EventAcronym{}
\EventYear{}
\EventDate{}
\EventLocation{}
\EventLogo{}
\SeriesVolume{}
\ArticleNo{}

\begin{document}

\maketitle

\begin{abstract}
\textbf{Background.}
    Requirements inspection (RI) is a well-established practice for detecting potential defects in requirements artifacts early in the software lifecycle. 
    Recent advances in large language models (LLMs) have stimulated 
    interest in their potential to support requirements engineering (RE)  tasks.
    However, empirical evidence on the effects of LLMs when used as collaborative assistants in human-performed RI remains scarce.
    \textbf{Aims.} 
    We aim to investigate the impact of LLM support on human-performed RI, considering inspection effectiveness in terms of smell identification and severity classification (i.e., nocuous \textit{vs} innocuous), as well as inspection duration.
    \textbf{Method.} 
    We conducted a controlled crossover design experiment with 34 participants,  
    who inspected textual specifications with and without LLM support, identifying and classifying requirements smells while recording inspection time.
    We analyzed the data using one Bayesian regression model per outcome variable, accounting for validity threats induced by the crossover design as well as covariates and mediators.
    \textbf{Results.} 
    Results show that LLM support negatively affects smell detection accuracy but has no significant effect on smell classification or task duration. 
    A learning effect is present across experimental periods, but reduced when RI is first performed with LLM support. 
    \textbf{Conclusions.} 
    Our findings provide empirical evidence that LLM support does not necessarily improve performance and may, instead, hinder it for novice inspectors.
    Moreover, the results suggest that learning RI with LLM-support from the beginning may slow down the skill acquisition process, implying threats for LLM-supported learning. 
\end{abstract}

\section{Introduction}
Software inspection is a well-established quality assurance practice that involves the systematic examination of software artifacts with the objective of identifying defects early in the development lifecycle. Since its formalization in the mid-1980s, inspection has been recognized as an effective and cost-efficient technique for improving software quality, reducing overall development costs~\cite{Fagan1986AdvancesIS,fagan2011history}. 
Inspections can be applied before executable artifacts exist, making them particularly valuable for detecting defects in requirements artifacts during the requirements engineering (RE) phase. Requirements inspection (RI)---a specialized form of inspection focused on these artifacts---plays a critical role in ensuring the quality, as defects introduced at the RE stage tend to propagate downstream and become expensive to fix~\cite{fernandez2017naming}.
In this context, the notion of \emph{requirements smells} has gained increasing attention. Requirements smells are indicators of potential quality issues---such as ambiguity, vagueness, or unverifiability---that may hinder understanding or lead to incorrect interpretations~\cite{Femmer17}. Although not all smells correspond to actual defects, they serve as early warning signals of violations of desirable quality properties. Distinguishing between nocuous (i.e., harmful) and innocuous smells is therefore essential, as it allows inspectors to prioritize issues more likely to negatively affect downstream activities~\cite{chantree2006identifying}. 

Recent advances in generative AI and large language models (LLMs) have stimulated increasing interest in their use to support RE activities. Prior work has investigated LLM-based approaches for analyzing and transforming requirements artifacts, including applications to code generation~\cite{Liu2022}, test scenario generation~\cite{Arora24,wang2025requirements}, and the derivation of UML sequence diagrams from natural-language requirements~\cite{ferrari2024model}, among others. More recently, Vogelsang et al.~\cite{vogelsang2025impact} investigated the impact of requirements smells on LLM performance in requirements-to-code traceability, further highlighting the sensitivity of LLM-based techniques to requirements quality.
%
However, existing studies primarily evaluate model performance and output quality~\cite{wolf2025quality}. In contrast, empirical evidence on how human-LLM collaboration
, where LLMs serve as supportive assistants in human-performed inspection, affects inspection performance remains largely unexplored. This gap is particularly critical for novice requirements analysts, who risk overreliance on LLMs, potentially hindering skill development. At the same time, organizations are increasingly adopting AI-assisted onboarding to compensate for the limited availability of senior experts for traditional mentoring\footnote{\url{https://newsletter.getdx.com/p/ai-cuts-developer-onboarding-time-in-half}. Visited 22 Jan 2026.}.


To address this gap, we conducted a controlled crossover experiment~\cite{vegas2015crossover}  in which students acting as novice inspectors performed RI tasks with and without ChatGPT (GPT-4o/GPT-4.1 models). Treating students as novice requirements analysts is common in empirical RE studies~\cite{BanoZ0S20,bano2018learning}. 
ChatGPT was selected among the available LLMs, given preliminary evidence of its potential~\cite{fantechi2023rule}. Our study focuses on the identification and severity classification of requirements smells (i.e., nocuous vs. innocuous), as well as on inspection duration. 


Our results indicate that using ChatGPT as a support tool affects requirements smell detection negatively, while it has no significant effect on smell severity classification. Moreover, inspection performance is influenced by a learning effect: participants generally perform better when executing the task for the second time. However, when the inspection task is first performed with ChatGPT support, this learning effect is reduced across 
periods. We also observe that longer inspection times are unexpectedly associated with poorer performance. 
Finally, smell detection and classification performance are positively correlated, indicating that participants who identify more smells also tend to classify them more accurately.

The main contributions of this work are:
\begin{itemize}
    \item A novel empirical investigation of the impact of using LLMs for RI.
    \item Evidence of learning and order effects in LLM-supported RI by novice inspectors.
    \item Practical insights into how (not) to use LLMs for RI by novice analysts.
\end{itemize}

The remainder of this paper is structured as follows. Section \ref{sect:backANDRel} introduces background concepts on RI, requirements smells, and the classification of nocuous and innocuous smells, and reviews related work. Section \ref{sect:studyDesign} describes the experimental design. Section \ref{sect:results} presents the results, while Section \ref{sect:discussion} discusses the findings. Section \ref{sec:threats} discusses threats to validity. Finally, Section \ref{sect:conclusion} concludes the paper and outlines directions for future work.


\section{Background and Related Work}\label{sect:backANDRel}

\subsection{Requirements Inspection and Smells}
Software inspection comprises a family of structured review techniques aimed at identifying violations of quality attributes in software artefacts~\cite{Fagan1986AdvancesIS}. When applied to RE, inspection focuses on systematically examining requirements artifacts to identify defects, inconsistencies, and indicators of poor quality that may compromise downstream development activities. 
Detecting defects in requirements artifacts early in the software lifecycle is critical to preventing their propagation into later development phases, where they can cause rework, delays, and quality issues in downstream activities, and where their correction becomes increasingly difficult and costly~\cite{fernandez2017naming,shull2000perspective}.

In this context, \textit{requirements smells} have been introduced as observable indicators of potential quality issues in requirements artifacts~\cite{Femmer17}. Requirements smells do not necessarily correspond to defects themselves, but signal situations that may hinder understanding, introduce ambiguity, or increase the likelihood of misinterpretation. As such, smells provide a practical means to guide inspection activities by drawing attention to requirements that warrant closer scrutiny.
In 2022, a catalogue of 206 requirements quality factors (i.e., smells) was published~\cite{Frattini2022}, derived from a systematic mapping study of 105 relevant papers~\cite{Montgomery2022}. Frattini et al. categorized these smells into three main classes: (i) \emph{lexical smells}, related to words or terms that may be misinterpreted; (ii) \emph{syntactic smells}, with problematic word(s) or sentence structures; and (iii) \emph{semantic smells}, which concern the interpretation of requirements within their usage context.

Among smells, it is important to distinguish between benign issues and those likely to cause misunderstandings and defects. 
In the early 2000s, Chantree et al. introduced the distinction between \textit{nocuous} and \textit{innocuous} phenomena~\cite{chantree2006identifying}. 
Nocuous issues are likely to result in divergent stakeholder interpretations 
and, consequently, incorrect or inconsistent implementations. In contrast, innocuous issues may exhibit potentially problematic characteristics at a superficial level, yet are typically interpreted consistently 
and are therefore unlikely to cause misunderstandings.
Accordingly, while smells indicate potential quality problems, not all necessarily require corrective action. 
Some may be easily disambiguated through 
context and domain knowledge and can therefore be considered innocuous~\cite{krisch2015myth}. Others 
may give rise to multiple plausible interpretations and thus represent a nocuous risk if left unaddressed~\cite{frattini2025applying}.
Distinguishing between nocuous and innocuous smells allows inspectors to prioritize issues more likely to negatively affect downstream development activities.

\subsection{Related Work}

Recent research has increasingly investigated the use of LLMs to support requirements quality assurance.
A systematic literature review by Wolf et al. synthesizes 26 peer-reviewed studies (2019–2025) on AI-based requirements quality assessment, highlighting a shift since 2023 toward LLM-driven approaches and semantically richer assessments~\cite{wolf2025quality}. 
Part of the works treat LLM as quality defect detectors~\cite{fantechi2023inconsistency,mahbub2024can,raj2024enhancing}, while others partially take into account also the human perspective, and the implications of using LLM for inspection~\cite{boukhlif2024using,bashir2025requirements,seifert2024can,lubos2024leveraging}.



\textit{LLM as Detectors.} In the first group, Fantechi et al.~\cite{fantechi2023inconsistency} explored the use of GPT-3.5 for detecting inconsistencies in natural-language requirements. Their study evaluated the LLM’s predictions against reference assessments provided by students. The results show that the model can identify certain inconsistencies, providing evidence of feasibility, while also highlighting variability in detection accuracy depending on inconsistency type  and requirements formulation.

Mahbub et al.~\cite{mahbub2024can} evaluated GPT-4’s performance in detecting multiple classes of defects in requirements artifacts, including ambiguities, inconsistencies, and incompleteness. Their study quantitatively assessed the model's precision across these defect types and different requirements versions. The results indicate that GPT-4 performs better in identifying incomplete requirements than in detecting inconsistencies and ambiguities. 
The authors 
concluded that, despite promising results, LLMs cannot replace human analysts in zero-shot settings due to limitations in domain-specific reasoning.

Raj et al.~\cite{raj2024enhancing} proposed a framework that leverages LLMs to identify, categorize, and resolve ambiguities in software requirements artifacts. 
The evaluation focuses on the LLM’s ability to identify ambiguity types and generate alternative formulations on a small evaluation set, providing preliminary evidence of LLMs' potential to support ambiguity-related quality improvement.

\textit{LLM and the Human Perspective.} In this group, Lubos et al.~\cite{lubos2024leveraging} investigated the use of an LLM to evaluate software requirements quality according to the ISO/IEC/IEEE 29148 standard. Their approach employs the LLM to assess quality characteristics, explain its judgments, and propose improved versions of requirements. The approach was assessed with the involvement of software engineers, who examined the usefulness of the LLM-supported analysis. The results suggest that LLMs can provide valuable support for requirements quality assessment and improvement, particularly through explanations and suggested refinements. 

Bashir et al.~\cite{bashir2025requirements} empirically studied the use of LLMs for ambiguity detection and explanation in industrial requirements. The authors evaluated performance across three industrial datasets and reported a 20.2\% average improvement in classification performance when using ten-shot prompting compared to zero-shot settings. In addition, they conducted human evaluations with industry experts, who assessed the generated explanations across dimensions including naturalness, adequacy, usefulness, and relevance, with an average score of 3.84/5.

Boukhlif et al.~\cite{boukhlif2024using} examined the use of LLMs to identify ambiguities, inconsistencies, and gaps in software requirements, with a particular focus on implications for software testing. The study compared LLM-based analysis with conventional SRS analysis techniques using quantitative metrics, as well as qualitative assessment of generated insights. The results indicate that LLMs can improve the accuracy and efficiency of requirements analysis, particularly in detecting ambiguous terms and uncovering inconsistencies that traditional approaches may overlook. 

Finally, Seifert et al.~\cite{seifert2024can} report a replication of an inspection experiment on requirements to assess whether LLMs can compete with human reviewers. They evaluate GPT-4-Turbo as the LLM reviewer. While the results indicate that LLMs can identify certain issues, the authors note that comparability with the original human inspection setting is constrained by necessary adaptations (e.g., how the LLM is trained and the material provided), and thus the results should be interpreted with care.


\textit{Contributions.} Overall, existing work provides valuable empirical evidence on the potential of LLMs to identify and reason about quality issues in textual requirements. However, prior research largely centers on evaluating LLM outputs---often with human involvement limited to assessing the correctness, usefulness, or plausibility of model-generated results---rather than examining their role within human-centered inspection processes. 
Human-in-the-loop scenarios are, however, closer to realistic industrial practice, where LLMs are increasingly viewed as support tools that assist rather than fully replace human reasoning and decision making in software engineering activities~\cite{natarajan2025human,takerngsaksiri2025human}. This is particularly important in mission-critical contexts, where requirements quality and human oversight remain crucial.
Furthermore, the use of chat-based LLMs is inherently iterative and dialogic, enabling a collaborative sensemaking process in which humans refine prompts, assess suggestions, and apply domain knowledge to inspection decisions.
%
Unlike prior work, this study adopts this realistic human-in-the-loop perspective by evaluating inspector performance when using an LLM as a support tool during RI, focusing on effectiveness, learning and order effects, and inspection duration. 
By adopting this perspective, our work complements existing research and addresses an important gap in empirical evidence on human–AI collaboration in RI.


\section{Study Design}\label{sect:studyDesign}
\label{sec:design}
The experimental design and procedures are described in accordance with the guidelines reported by Jedlitschka et al.~\cite{jedlitschka2008reporting} and Vegas et al.~\cite{vegas2015crossover}.

\subsection{Goal and Research Questions}

The overall goal of the experiment is as follows:
\smallskip

\hspace*{-5mm}
\begin{tikzpicture}
    \node [mybox] (box){%
        \begin{minipage}{.95\columnwidth}
        \centering
        Understanding the \textbf{potential of using LLMs to support 
        RI
        } for novice inspectors, and the \textbf{impact of LLM-use on their learning behavior}.
        \end{minipage}
    };
\end{tikzpicture}

To achieve the goal, our aim is to answer the following research questions (RQs):
\begin{description}
    \item[RQ1] What is the impact of LLM support on the time required to perform RI? 
    \item[RQ2] What is the impact of LLM support on the accuracy of requirements smell identification?
    \item[RQ3] What is the impact of LLM support on the accuracy of classifying identified requirements smells as nocuous? 
\end{description}

\subsection{Selection of Subjects}
Participants were BSc’s students in Computer Engineering at the University of Florence, enrolled in the Industrial Computer Science course. They were opportunistically recruited based on availability. Participation was voluntary and granted a bonus toward the course exam; however, students could achieve high exam scores regardless of participation through the standard examination process. Prior to the experiment, all participants were informed about the study goals and procedures and provided written informed consent. They were explicitly informed that the collected data would be used exclusively for research purposes, analysed in an anonymous and aggregated form, and that no link between their identity and the recorded responses would be retained.


Participants were divided into two experimental groups (group 1 and group 2, both containing 17 participants each).
When allocating participants to groups, we blocked on the factors \emph{proficiency in RI} and \emph{proficiency in using LLMs}, collected during a pre-test questionnaire (cf. Section~\ref{sect:expObj}) on a five-point ordinal scale.
They were assigned to groups using optimal matching based on Mahalanobis distance over these two variables.
Ultimately, both experimental groups contained a similiar number of participants (with a maximum difference of 2) for each of the scales' five levels of both factors.
This way, the effect of these factors is evenly dispersed among the groups and does not confound the effect of interest.

\subsection{Experimental Objects and Tasks}\label{sect:expObj}
The experiment was conducted over two sessions, during which participants performed RI tasks.
To perform the tasks, participants were provided with (i) a requirements document, (ii) a reference document providing definitions and examples of the selected smell types and the nocuous/innocuous distinction, and (iii) a test document to support RI. 
Participants were asked to inspect each requirement in the provided requirements document and determine whether it exhibited one of the predefined smell types; classify identified smells as nocuous or innocuous; record the start and end times of the inspection task.
Additionally, participants were provided with (iv) a pre-test questionnaire before the inspection sessions, and (v) post-test questionnaires after each inspection session.

\begin{table}
\scriptsize
    \centering
    \caption{Types of considered smells}
    \label{tab:smells}\vspace*{-0.25cm}
    \begin{tabularx}{\columnwidth}{@{}lX@{}}
    \toprule
     \textbf{Smell categories and types}& \textbf{Description}\\
    \midrule
    \textbf{Lexical smells} & \\
\hspace{1em} subjective language & Words of which the semantics are not objectively defined (e.g., \textit{user-friendly}, \textit{easy to use}, \textit{cost effective}). 
Sentences expressing personal opinions or feelings.\\
\hspace{1em} optional parts &Sentences containing optional parts, e.g., by using words such as \textit{possibly}, \textit{eventually}, \textit{if possible}, \textit{if needed}, etc.\\
\hspace{1em} weak verbs &Weak verbs, such as \textit{can}, \textit{could}, \textit{may}, etc.\\ [1ex]
    \textbf{Syntactic smells} & \\
\hspace{1em} vague pronouns &Pronouns that refer back to a previous part of the text for which the reference is unclear.\\
\hspace{1em} passive voice &Sentences using passive voice such that it is unclear who is performing a certain action.\\
\hspace{1em} negative phrases &Sentences containing negative expressions or modifiers (e.g., \textit{not}).\\ [1ex]
    \textbf{Semantic smells} &\\
\hspace{1em} logical inconsistencies& Two requirements, which are connected to the same concepts, contradicting each other.\\
\hspace{1em} numerical discrepancies & Two requirements connected to the same concepts, containing inconsistent and/or contradicting numerical information. \\
\hspace{1em} ambiguities & Unclear/imprecise sentence parts that can be misunderstood if read by different people.\\
    \bottomrule
    \end{tabularx}
    \vspace{-1em}
\end{table}

\textbf{(i) Requirements document.}
The documents used in the two sessions contained a set of requirements for two distinct video game implementations: the Arkanoid game in the first session and the Snake game in the second session. These documents 
contain smelly and non-smelly requirements and were designed to be comparable in terms of size and number of smells (i.e., Arkanoid: 40 requirements, 21 smells; Snake: 39 requirements, 19 smells). 
The smells contained cover the nine smell types from~\cite{Frattini2022} listed in \Cref{tab:smells}. We also classified each smell as nocuous or innocuous.


The requirements documents were extended versions of two documents previously used in~\cite{vogelsang2025impact}. Starting from these original documents, one of the authors extended the documents and injected additional smells to ensure coverage of all nine selected smell types, with each smell type appearing at least once across the documents and each requirement containing at most one smell\footnote{The choice of constraining each requirement to contain at most one smell, although not fully representative of real-world requirements documents, was made to simplify both the inspection task for participants and the subsequent analysis. In particular, allowing multiple smells per requirement would have increased annotation ambiguity for participants and complicated the computation and interpretation of performance measures (e.g., precision and recall).}. The same author also performed an initial nocuous/innocuous classification of the injected smells.
Subsequently, two additional authors independently inspected the requirements documents, identifying the presence of smells and classifying them as nocuous or innocuous. The three authors then jointly reviewed all identified discrepancies through an iterative, consensus-oriented discussion process until a shared decision was reached for each requirement.
Given the inherently subjective and context-dependent nature of requirements smells and their perceived severity~\cite{gervasi2019}, we adopted a dialogic reconciliation process rather than relying on formal inter-rater agreement measures. This process was applied consistently to both smell identification and nocuous/innocuous classification. Based on the final agreed-upon version of the documents, a reference annotation was established for each requirement, specifying the presence or absence of a smell, the smell type, and its nocuous or innocuous classification. This annotated version of the documents served as the ground truth for the subsequent analysis of participants’ inspection performance. Although the initial annotation from one of the authors could have biased the ground truth, we argue that the multiple iterations and re-evaluations mitigate this possible bias.


\textbf{(ii) Reference document.} Participants received a reference document containing the definition of each smell type, along with two illustrative examples and their corresponding explanations. The document also included the definition of the nocuous and innocuous classes, together with two examples for each class.

\begin{figure}
    \centering
    \includegraphics[width=\linewidth]{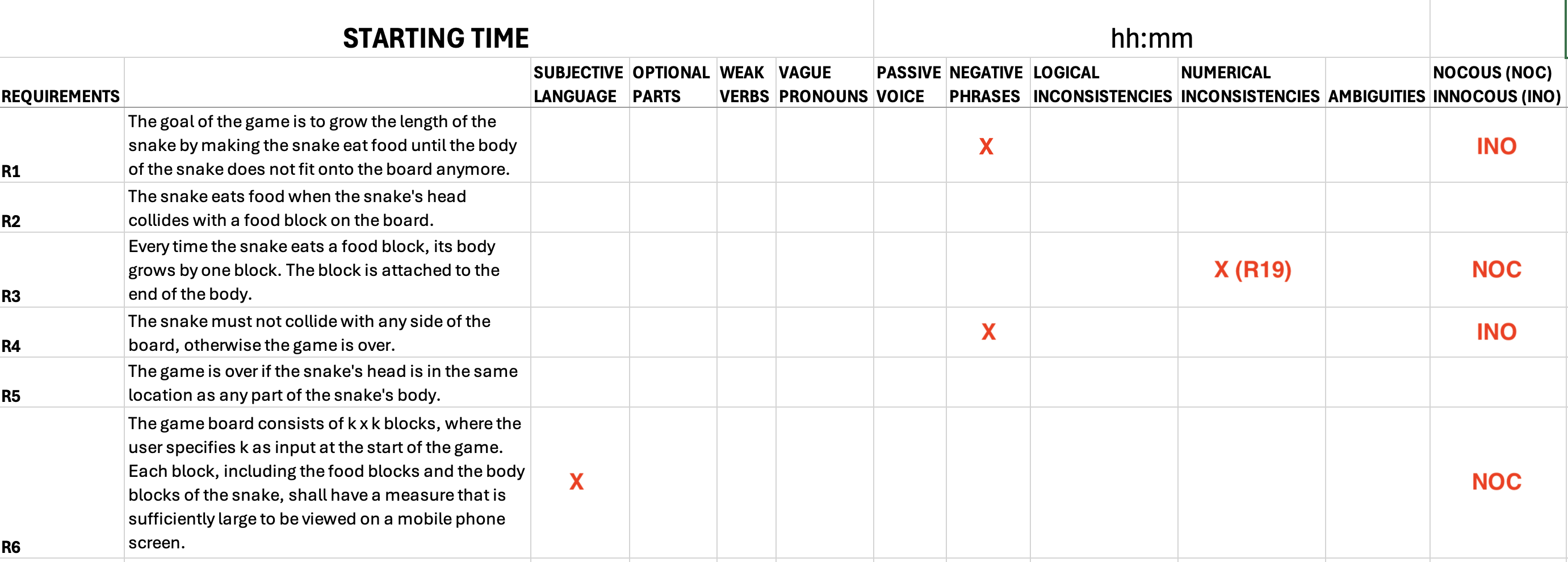}
    \caption{Excerpt of the test document. Red annotations indicate example expected completions and were not shown to participants. }
    \label{fig:testDoc}
\end{figure}

\textbf{(iii) Test document.} Participants received a spreadsheet to support the inspection task (an excerpt of which is shown in Fig. \ref{fig:testDoc}). Each row corresponded to a single requirement, while columns represented the nine selected smell types, allowing participants to indicate whether a given requirement exhibited a specific smell. An additional field was provided to classify the identified smell as nocuous or innocuous. The spreadsheet also included dedicated fields for recording the start and end time of the inspection task.

\textbf{(iv) Pre-test questionnaire.}
Before the inspection task, participants completed a pre-test questionnaire collecting demographic and background information. The questionnaire gathered the identifier, gender, age, and the highest degree completed by each participant. In addition, participants self-reported their proficiency in RI and in using LLMs using a five-point scale (1--5). For participants who reported proficiency in RI greater than 1, the questionnaire also collected information about the types of RI tasks with which they were familiar and the contexts in which they had previously performed such tasks.

\textbf{(v) Post-test questionnaire.}
After each inspection session, participants completed a post-test questionnaire tailored to the experimental condition they experienced, distinguishing between LLM-supported and unsupported inspections and between period~1 and period~2.
Collectively, across all sessions and experimental conditions, the post-test questionnaires gathered qualitative comments and feedback on the inspection task, participants’ preferences between the two variants (LLM-supported vs. non-supported), and self-reported proficiency in English. When LLM support was used, participants reported the LLM version and the number of requirements inspected with LLM assistance.



\subsection{Choice of Design and Variables}\label{sec:design:variables}
To achieve the objectives of the study, we conducted a crossover experiment~\cite{vegas2015crossover}. The experiment was carried out over two periods (period~1 and period~2), separated by a one-week washout interval. In period~1, group~1 performed the RI task without the support of ChatGPT, while group~2 performed the same task with ChatGPT support; in period~2, the exposure conditions were reversed. During the study (May 2025), participants used ChatGPT models from the GPT-4o and GPT-4.1 families, with model availability depending on subscription tier and quota limitations. Each period involved the inspection of a different requirements document (cf.\ Section~\ref{sect:expObj}).



\Cref{tab:vars} lists all variables involved in the study.
The \textbf{Type}-column differentiates the main factor of interest (exposure), the three dependent variables (outcome), the factors induced by the crossover-design (design), and additional factors that may affect the outcomes (covariate).

\begin{table}
    \caption{List of variables involved in the study}
    \label{tab:vars}
    \footnotesize
    \centering
    \begin{tabularx}{\linewidth}{lp{2cm}lXp{2.5cm}}
        \toprule
        \textbf{Variable} & \textbf{Name} & \textbf{Type} & \textbf{Description} & \textbf{Scale} \\
        \midrule
        \textit{llm} & LLM-Support & exposure & Whether or not an LLM was used during RI & \{gpt, nogpt\} \\
        \hline
        \textit{duration} & Task Duration & outcome & Number of minutes taken to complete the task & $\mathbb{N}_0$ \\
        \textit{detection} & Detection Accuracy & outcome & Macro-averaged $F_1$-score computed across all smell types for smell detection
        & $[0; 1]$ \\
        \textit{classification} & Classification Accuracy & outcome & $F_1$-score of correctly classifying requirements smells as nocuous, computed only on correctly detected smells
        & $[0; 1]$ 
        \\
        \hline
        \textit{period} & Experimental Period & design & Time slot of the observation & \{period1, period2\} \\
        \textit{sequence} & Treatment Sequence & design & Order in which the participants performed the task with/without LLM-support & \{group1, group2\} \\
        \textit{subject} & Participant ID & design & ID of the participant & 1--34 \\
        \hline
        \textit{prof.llm} & Proficiency using LLMs & covariate & Self-reported proficiency in using LLMs & \{1, 2, 3, 4, 5\} \\
        \textit{prof.ri} & Proficiency in RI & covariate & Self-reported proficiency in performing RI tasks & \{1, 2, 3, 4, 5\} \\
        \textit{prof.english} & Proficiency in English & covariate & Self-reported proficiency in the English language & \{1, 2, 3, 4, 5\} \\
        \textit{gpt.use} & GPT Usage & covariate & Self-reported number of requirements for which the LLM was used during the inspection task (when available)
        & \{0, 1--10, 11--20, 21--35, 36--40\} \\
        \bottomrule
    \end{tabularx}
\end{table}

\subsection{Experimental Procedure}
\subsubsection{Experimental Phases} The experiment began with a training phase in which participants attended an introductory lesson designed to provide all the information necessary to perform the inspection task successfully. The lesson covered the following topics: RI, requirements smells, the distinction between nocuous and innocuous smells, and basic principles of prompt engineering. To increase engagement and motivation, the lesson incorporated gamification elements by framing the activity as a simulated hiring scenario in which participants were employees of a video game company competing for the ``employee of the month'' distinction.
Following the introductory lesson, participants were asked to complete an online pre-test questionnaire no later than three days before the first inspection task session.
One week after the introductory lesson, participants entered the inspection phase, which constituted the core phase of the experiment. This phase was conducted over two sessions (period 1 and period 2), separated by a one-week washout interval. The inspection tasks were performed without time constraints. Participants initiated the inspection during an in-class session and were allowed to complete it outside the classroom if necessary, provided that they accurately recorded the start and end times of the task. Although this variability in working conditions reduces control, 
this choice was made to prioritize data completeness and avoid comparisons between fully
and partially completed inspections. Only two participants continued after the in-class session.
%
After each inspection task session, participants uploaded the completed test document to a shared online folder labeled with their participant ID and completed an online post-test questionnaire, which varied according to the experimental period and assigned condition. The pre-test and post-test questionnaires were administered through an online survey platform.


\subsubsection{Measurement Procedure}
To quantify participants' performance in identifying requirements smells, we calculated effectiveness metrics by comparing each participant's annotations with the ground truth.
Each participant completed the RI task by marking, for each requirement, the presence or absence of a predefined set of smell types. 

The comparison with the ground truth was performed at the level of individual smell types. Specifically, for each participant and for each smell type, we computed an $F_1$-score by treating the identification of that smell across all requirements as a binary classification problem. True positives, false positives, and false negatives were derived by aligning participant annotations and ground-truth labels based on requirement identifiers.
Participant-level performance was summarized by computing the macro-averaged $F_1$-score across smell types. This metric is common in the evaluation of RE tasks~\cite{berry2021empirical}. Although $F_\beta$ can give a more nuanced and practically relevant viewpoint~\cite{berry2021empirical}, the  computation of the actual value of $\beta$ would have required a dedicated experimental evaluation, and it is left for future work. 


Classification performance, again evaluated against the ground truth,  was computed  only for smells that were correctly identified, as classification presupposes successful detection; including undetected smells would conflate detection errors with classification errors and would therefore not reflect classification ability per se. For each participant, classification effectiveness was computed by treating the assignment of the nocuous label as a binary classification problem over the set of correctly identified smells. Specifically, true positives, false positives, and false negatives were derived by aligning participants’ nocuous classifications with the ground-truth nocuous labels. We focused on the nocuous class because correctly identifying nocuous smells is critical in practice, as these smells are more likely to lead to misunderstandings and downstream defects if left unaddressed. Participant-level classification performance was summarized using the $F_1$-score. The same considerations about $F_\beta$ discussed above apply also for this task.

\subsection{Data Analysis}
\label{sec:design:analysis}

To analyze the collected data, we utilized Bayesian data analysis (BDA) within the framework for statistical causal inference (SCI)~\cite{siebert2023applications,pearl2009causal}. 
BDA techniques are more complex to apply than their frequentist counterpart, but offer richer insights that are more intuitive to interpret and more robust against uncertainty~\cite{furia2019bayesian,frattini2024second}.
In our particular analysis, BDA allows us to properly model the outcome variables using a distribution family with maximum entropy~\cite{jaynes2003probability}.
Whereas frequentist methods differentiate only normally and non-normally distributed outcomes, BDA distinguishes many more distribution families~\cite{gren2021possible}.
Since none of our outcome variables follow a normal distribution, BDA is the the more appropriate choice~\cite{mcelreath2018statistical}.

The SCI framework ensures that identified effects are causal and not correlational, within the scope of the collected data~\cite{pearl2009causal}.
For the overarching SCI framework, we adhered to the three-step process proposed by Siebert~\cite{siebert2023applications}, which is based on the seminal work by Pearl~\cite{pearl2009causal}.
Despite SCI principles being particularly effective for analyzing observational data~\cite{furia2022applying}, they also support analyzing experimental data by distinguishing total from direct effects~\cite{frattini2025applying} and clearly guide the analysis in crossover-design experiments~\cite{vegas2015crossover,frattini2024crossover}.
For BDA, we followed established textbooks by McElreath~\cite{mcelreath2018statistical} and Gelman et al.~\cite{gelman1995bayesian}.

\subsubsection{Modeling}
\label{sec:design:analysis:modeling}

\begin{figure}
    \centering
    \includegraphics[width=.55\linewidth]{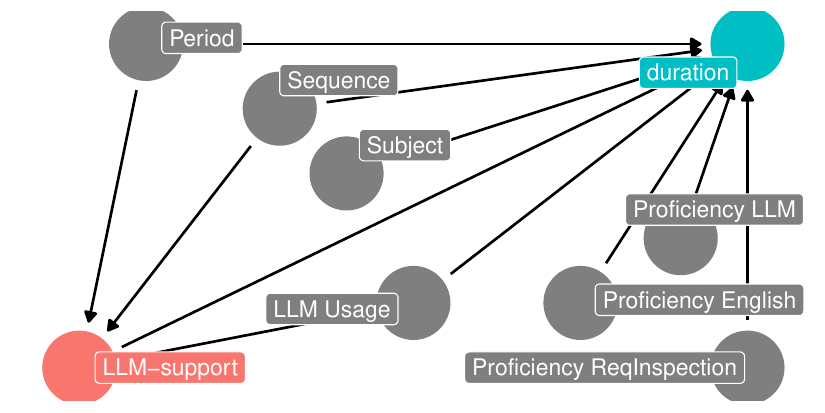}
    \caption{DAG for RQ1}
    \label{fig:dag:rq1}
\end{figure}

In the first step proposed by Siebert~\cite{siebert2023applications} titled \textit{modeling}, we explicitly encoded our causal assumptions in a directed, acyclic graph (DAG)~\cite{glymour2008causal}.
In a DAG, nodes represent relevant variables and directed edges assumed causal relationships between these variables.
For each RQ, we modeled one DAG centered around the relationship between the exposure (i.e., the use of ChatGPT) and the respective outcome variable (i.e., duration, detection performance, and classification performance).
\Cref{fig:dag:rq1} shows the DAG for RQ1 with the exposure colored red and the outcome cyan.
The DAGs for RQ2 and RQ3 differ only (1) by the respective outcome variable, and (2) by including \textit{duration} as a mediator variable.

\subsubsection{Identification}
\label{sec:design:analysis:identification}

Based on the DAG for each RQ, we performed the second step of the SCI framework~\cite{siebert2023applications} named \textit{identification}.
During this step, we identified all variables beyond the main exposure (LLM-support) that we need to statistically control in order to avoid any bias. 
To this end, we applied d-separation~\cite{pearl2009causal}, which identified the adjustment set of variables necessary to de-confound a causal effect of interest.
In the exemplary DAG for RQ1 in \Cref{fig:dag:rq1}, the variables \textit{period} and \textit{sequence} confound the effect of \textit{LLM-support} on \textit{duration}, and hence, must be included in the adjustment set~\cite{pearl2009causal}.
The variable \textit{LLM-usage} acts as a mediator, which we included to isolate the direct effect of interest~\cite{mcelreath2018statistical}.
Finally, the remaining variables act as simple covariates, which we included to improve the precision of the effect estimation~\cite{cinelli2024crash}.

\subsubsection{Estimation}
\label{sec:design:analysis:estimation}

With the adjustment set determined for each RQ, we conducted the final \textit{estimation} step~\cite{siebert2023applications} to estimate the direction and strength of the effect that the statistically controlled variables have on the outcome. 
For this process, we followed common guidelines for BDA~\cite{mcelreath2018statistical,gelman1995bayesian}.

We started by selecting a distribution family with maximum entropy for each of the three outcome variables~\cite{jaynes2003probability}:
We treated every outcome variable as a random variable and selected the least restrictive distribution family that is consistent with all epistemological and ontological assumptions~\cite{mcelreath2018statistical}.
This yielded two different distribution families:

\begin{itemize}
    \item The outcome variable \textit{duration} was modelled as a \textbf{negative binomial} (or: gamma Poisson) distribution as the variable represents a discrete count (in number of minutes) with overdispersion ($\textit{VMR} = 7.2 > 1$).
    \item The outcome variables \textit{detection performance} and \textit{classification performance} were both modeled as \textbf{beta} distributions since they are measured as $F_1$-scores bounded between 0 and 1. Given the existence of 0.0 and 1.0 scores occurring in both variables, we use the \textbf{zero-one-inflated beta} distribution variant specifically.
\end{itemize}

Next, we designed a Bayesian regression model per RQ.
We specified the shape-defining parameter of each distribution with a linear model containing all variables from the adjustment set.
For example, the regression model for RQ1 is defined as follows:
\begin{align*}
    duration_i & \sim \text{NegBinom}(n_i, p_i) \\
    \text{logit}(p_i) & \leftarrow \alpha + \alpha_{\textit{ID}} + \beta_{\textit{llm}} \cdot {llm}_i + \beta_{\textit{llm} \times \textit{prof.llm}} \cdot {llm}_i \cdot \textit{prof.llm}_i + \dots
\end{align*}

The $\dots$ represent the remaining factors included in the model, i.e., all of the remaining factors listed in \Cref{tab:vars} and shown in \Cref{fig:dag:rq1}.
For brevity of the presentation, we obscured them in the manuscript.
The regression model $\text{logit}(p_i) \leftarrow \dots$ consists of several components:

\begin{itemize}
    \item A population-level intercept $\alpha$ representing the general difficulty of the experimental task
    \item A group-level intercept $\alpha_{\textit{ID}}$ representing the general skill of a participant. This models subject-level variability~\cite{vegas2015crossover}.
    \item Population-level effects $\beta_X \cdot X_i$, where $X_i$ is the value of variable $X$ (e.g., the use of ChatGPT, labelled $\textit{llm}$) in observation $i$ and $\beta_X$ is a probability distribution representing the effect that these values have on the outcome variable.
    \item Interaction effects $\beta_{X \times Y} \text{X}_i \cdot \text{Y}_i$ where one variable $Y$ moderates the effect of another variable $X$ on the outcome.
\end{itemize}

Every intercept ($\alpha$ and $\alpha_{\textit{ID}}$) and effect coefficient (all of $\beta_X$ and $\beta_{X \times Y}$) is a probability distribution of the effect strength of the respective factor on the outcome variable.
To each of these, we assign a prior probability distribution, i.e., encode our assumptions of feasible effect strengths prior to observing the data~\cite{mcelreath2018statistical}.
Given the lack of strong empirical evidence for either of these factors, we use uninformative priors like $\mathcal{N}(\mu=0, \sigma=1)$~\cite{wesner2021choosing}, i.e., normal distributions centered around 0 with a wide variance 1.
This represents the belief that any factor could have either a positive ($\mu > 0$), negative ($\mu < 0$), or nonexistent ($\mu \approx 0$) effect on the outcome.
To assess the eligibility of the selected priors, we perform prior predictive checks~\cite{wesner2021choosing}.
In this process, we sample from the Bayesian regression model without updating the priors via the collected data. 
The results of this prior predictive check visually showed that the range of potential outcomes encompass the actual outcomes, confirming the plausibility of the priors.

With the priors' eligibility confirmed, we continued with model training.
We implemented the analysis using the \textsl{R} library \texttt{brms}~\cite{burkner2017brms} acting as an interface to the statistical modelling language \textsl{STAN}~\cite{carpenter2017stan}.
During the training process, Hamiltonian Monte Carlo Markov Chains (MCMC) updated the coefficient distributions based on the collected data~\cite{brooks2011handbook}.
This way, the parameters of the coefficient distributions (e.g., $\mu$ and $\sigma$) were adjusted to better reflect the outcome variable based on the predictor variable values.
After training, we assessed several diagnostics to ensure the validity of the training process.
The Gelman-Rubin convergence statistic showed acceptable values $\hat{R} \in ]0.99, 1.01[$ for all coefficients~\cite{gelman1992inference}.
Similarly, the proportion of effective sample sizes was above the threshold $\frac{n_{\textit{eff}}}{n} > 0.1$ for all coefficients~\cite{mcelreath2018statistical}.
We observed no divergent transitions, raising concerns about the training process.
Finally, we performed posterior predictive checks by drawing samples from the now-updated coefficient distributions, which confirmed that the training process moved the estimates closer and tighter to the observed values.
Note that, since our goal is not out-of-sample prediction but rather causal inference, we do not perform train-test splits on the available data.

To obtain interpretable results on the outcome space, we calculated marginal and conditional effects and plotted their distribution.
Marginal and conditional effects are estimated from the model by varying only one or two factors while holding all others constant at representative levels.
This way, the isolated effect of that factor on the outcome variable becomes clear, showing how it affects the outcome variable.

\section{Study Results}
\label{sect:results}

The following sections present the results obtained from the analysis about whether the use of ChatGPT in RI 
impacts the duration of the task (RQ1 in \Cref{sec:results:rq1}), the accuracy of detecting smells (RQ2 in \Cref{sec:results:rq2}), and the accuracy of correctly classifying correctly detected smells as nocuous (RQ3 in \Cref{sec:results:rq3}). 
Due to the page limitation, this section reports only meaningful results.
All descriptive statistics, DAGs, analyses, results, and figures can be found in our replication package.

\subsection{RQ1: Impact on Task Duration}
\label{sec:results:rq1}

\begin{figure}
     \centering
     \begin{subfigure}[b]{0.48\textwidth}
         \centering
         \includegraphics[width=\textwidth]{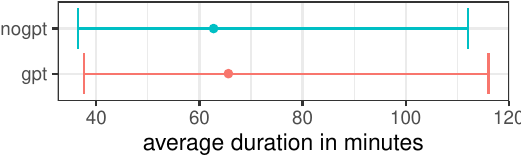}
         \caption{Effect on task duration}
         \label{fig:result:rq1:llm}
     \end{subfigure}
     \hfill
     \begin{subfigure}[b]{0.48\textwidth}
         \centering
         \includegraphics[width=\textwidth]{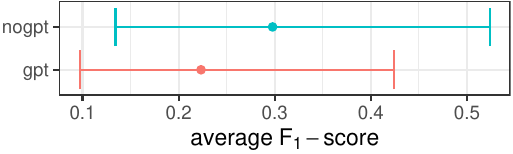}
         \caption{Effect on detection accuracy}
         \label{fig:result:rq2:llm}
     \end{subfigure}
        \caption{Marginal effects of LLM-support}
        \label{fig:result}
\end{figure}

\Cref{fig:result:rq1:llm} shows the marginal effect of LLM-support on the task duration.
The mean duration values (points in \Cref{fig:result:rq1:llm}) are close, and the 95\%-credibility intervals (whiskers in \Cref{fig:result:rq1:llm}) overlap, indicating that the use of ChatGPT does not significantly change the task duration.

Apart from that, no other predictor in the linear model for task duration showed a significant effect. 
At most, the \textit{period} variable shows a slight decrease in duration in Period 2 (by about 5 minutes).
This decrease is not significant but expected as an maturation effect observed in crossover-design experiments~\cite{vegas2015crossover}.

\subsection{RQ2: Impact on Detection Accuracy}
\label{sec:results:rq2}

\Cref{fig:result:rq2:llm} shows the marginal effect of LLM-support on the accuracy of detecting requirements smells.
Observations from the control group (i.e., participants not using ChatGPT) achieved about 8\% higher $F_1$-scores on average.
The total uncertainty around these averages, represented by the 95\%-credibility intervals, still overlaps.

\begin{figure}
     \centering
     \begin{subfigure}[t]{0.55\textwidth}
         \centering
         \includegraphics[width=\textwidth]{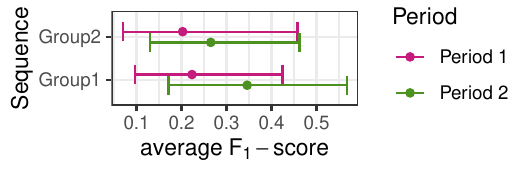}
         \caption{Conditional effect of the experimental period and sequence}
         \label{fig:result:rq2:carryover}
     \end{subfigure}
     \hfill
     \begin{subfigure}[t]{0.40\textwidth}
         \centering
         \includegraphics[width=\textwidth]{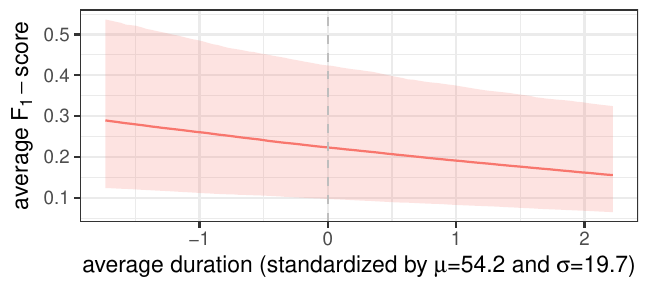}
         \caption{Marginal effect of the duration}
         \label{fig:result:rq2:duration}
     \end{subfigure}
     
     \caption{Marginal effects of design variables on detection accuracy}
     \label{fig:result:rq2}
\end{figure}

\Cref{fig:result:rq2:carryover} visualizes the conditional effect of both \textit{period} and \textit{sequence} on the detection accuracy.
Observations in the second period achieved better $F_1$-scores by more than 10\% in both sequences.
This clearly supports a maturation effect typical for crossover-design experiments~\cite{vegas2015crossover,frattini2024crossover}.
Across both periods, group 1 (who performed the task without ChatGPT in period 1) reached slightly higher accuracy, suggesting an optimal sequence~\cite{vegas2015crossover}.
Significantly though, the conditional effect of both \textit{period} and \textit{sequence} shows a clear effect.
Participants in group~1, who started the inspection task without ChatGPT support, exhibited a greater maturation effect---a $\sim12\%$ improvement in performance between periods~1 and 2---compared to participants in group~2, who started with ChatGPT support and improved by only $\sim7\%$.
\Cref{fig:result:rq2:carryover} shows this through a greater distance between the accuracy average between the two periods in group 1 than group 2.
This is a clear indication of a carryover effect possible in crossover-design experiments~\cite{vegas2015crossover}.



Additionally, longer task durations are associated with lower $F_1$-scores, as seen in \cref{fig:result:rq2:duration}.
None of the other covariates show an effect on the detection accuracy.


\subsection{RQ3: Impact on Classification Accuracy}
\label{sec:results:rq3}

\begin{figure}
     \centering
     \begin{subfigure}[t]{0.48\textwidth}
         \centering
         \includegraphics[width=\textwidth]{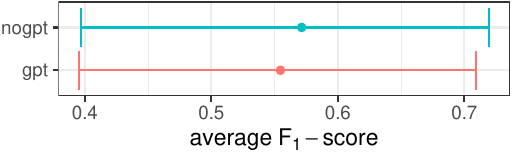}
         \caption{Marginal effect of LLM-usage}
         \label{fig:result:rq3:llm}
     \end{subfigure}
     \hfill
     \begin{subfigure}[t]{0.48\textwidth}
         \centering
         \includegraphics[width=\textwidth]{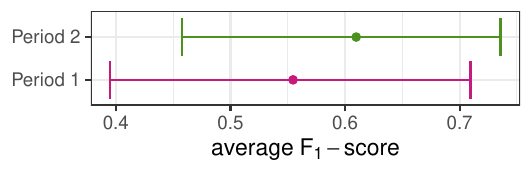}
         \caption{Marginal effect of the experimental period}
         \label{fig:result:rq3:period}
     \end{subfigure}
        \caption{Marginal effects of design variables on classification accuracy}
        \label{fig:result:rq3}
\end{figure}

\Cref{fig:result:rq3:llm} shows that using ChatGPT has virtually no effect on accuracy when classifying correctly detected smells as nocuous.
Again, the classification accuracy also improves between the periods as visualized in \Cref{fig:result:rq3:period}, though this effect is less strong.
Similarly, there is a notable yet weaker carryover effect where the $F_1$-score of classifying smells in group 1 (starting without ChatGPT) improves by about 5\% between periods and in group 2 only by about 3\%. 
Also, longer task durations produce worse classification accuracy.
All other covariates do not show any significant effect.

\begin{highlightbox}{\faLightbulb~Answer to RQs}
    When using LLMs for RI
    , participants took equally long time (RQ1) but detected smells less precisely with $\Delta F_1 \approx -8\%$ (RQ2). 
    Notably, performing the task first with and later without LLMs reduced the learning effect (i.e., the maturation-based improvement of $F_1$-scores between periods) by half (6\% instead of 12\% improvement).
    The severity classification accuracy remains unaffected by LLM usage (RQ3).
\end{highlightbox}
\section{Discussion}\label{sect:discussion}
Contrary to the common expectation that LLMs may uniformly improve performance in RE tasks raised by  prior work studying LLMs in isolation~\cite{mahbub2024can,bashir2025requirements,boukhlif2024using}, the results highlight a more nuanced picture.
In particular, while ChatGPT support did not significantly affect inspection time or smell classification accuracy, it was associated with decrease in smell detection accuracy and reduced learning effect when inspection was first performed with LLM support. Based on these findings, we discuss relevant observations and implications.

\subsection{Key Observations}

\textit{Potential Over-Reliance and Automation Bias.} The negative effect of LLM support on smell detection accuracy suggests that, for novice inspectors, the availability of an LLM may interfere with careful, systematic requirements analysis. One plausible explanation is that participants may have delegated part of the inspection effort to the LLM, relying on its responses instead of engaging deeply with the requirements. This aligns with concerns raised in prior research about over-reliance on AI-generated suggestions and automation bias, particularly when users lack strong domain expertise~\cite{passi2022overreliance,carnat2024human}.
Similar concerns have been reported in qualitative studies of human–LLM collaboration in other software engineering activities. Hamza et al. observed that, during a hands-on workshop with professional software engineers using ChatGPT, participants sometimes exhibited social loafing and excessive reliance on the AI, reducing intellectual engagement and critical thinking~\cite{hamza2024human}. 

\textit{Distinct Cognitive Processes in Smell Detection and Severity Classification.} Interestingly, LLM support did not significantly influence smell severity classification accuracy once smells were correctly identified. This suggests that, when participants successfully detected a smell, they were generally able to reason about its severity (nocuous vs. innocuous) regardless of LLM support. 
A similar pattern was observed in prior work on RI, where inspectors provided accurate justifications once a violation had been correctly identified, independently of the supporting representation~\cite{broccia2026rolecognitiveabilitiesrequirements}. Classification may therefore rely more on conceptual understanding of quality risks than on the mechanical detection of smell indicators. This distinction underscores the importance of treating detection and classification as two cognitively distinct activities in RI.  

\textit{Reduced Skill Development with Early LLM Support.} The observed learning effect across experimental periods aligns with established methodological practice in crossover designs, which requires learning effects to be explicitly examined and accounted for~\cite{vegas2015crossover,frattini2024crossover}. Participants  performed better in the second inspection session, indicating that repeated exposure to the task led to improved effectiveness. However, the reduced learning effect observed when the inspection task was first performed with LLM support is particularly relevant. It suggests that early reliance on LLMs may hinder the development of inspection skills, potentially by short-circuiting the reflective processes through which novice inspectors internalize inspection strategies and quality heuristics.

\textit{Longer Duration Does Not Imply Better Performance.} The negative association between longer inspection duration and performance indicates that spending more time on the task did not translate into higher accuracy. Rather than reflecting thoroughness, longer durations may signal uncertainty, lack of strategy, or difficulties in understanding the requirements. These difficulties may have been exacerbated by participants’ self-reported moderate English proficiency (average $3.5$ on a 1–5 scale), which could have hindered requirements comprehension. This further supports the interpretation that effective inspection depends more on structured reasoning than on time investment alone.

\subsection{Implications for Practice, Research, and Teaching}
\textit{Implication for Practice.} For practitioners, these results caution against uncritical adoption of LLMs as support tools in RI, especially for novice inspectors. While LLMs can generate plausible explanations and suggestions, their use may inadvertently reduce inspectors’ engagement with the requirements and weaken detection performance. Organizations introducing LLMs into inspection workflows should therefore consider restricting their use to specific phases (e.g., post-inspection review or justification refinement) rather than as a primary aid during initial defect detection. 
Similar concerns about the impact of early generative AI use on skill development have also been reported in software development contexts, where developers expressed worries that novices would become too reliant on generative AI tools, missing opportunities to develop coding and problem-solving skills~\cite{stray2025human}.

The lack of time savings suggests that LLM support does not necessarily lead to efficiency gains in inspection tasks. Practitioners should not assume that LLM-assisted inspection will be faster or more effective without careful evaluation in their specific context.
Similar patterns have been observed in other software engineering activities.
Stray et al.~\cite{stray2025developerproductivitygithubcopilot} report that the introduction of GitHub Copilot was not associated with substantial, consistently measurable productivity gains in development tasks, despite developers’ strong perception of increased efficiency. This gap between perceived and observed benefits reinforces our findings that LLM support should not be assumed to yield performance or efficiency improvements without careful, context-specific evaluation. 

On the other hand, LLMs may still support onboarding novice engineers when positioned as aids for learning and orientation rather than substitutes for human reasoning or task execution. Recent studies suggest that LLMs can help newcomers navigate documentation, reduce information overload, and explore unfamiliar technical material in a self-directed manner. However, effective onboarding still requires sustained human mentoring and supervision~\cite{azanza2024can,adejumo2024towards}.

\textit{Implications For Research.} From a research perspective, this study contributes empirical evidence to the still underexplored area of human--LLM collaboration in RI. The results highlight the need to move beyond evaluating LLM output quality in isolation and toward studying how LLMs shape human behavior, learning, and decision-making processes.
Future research should investigate alternative integration strategies, such as constrained or staged LLM usage, and examine whether different prompting styles, explanations, or transparency mechanisms mitigate the negative effects observed here. Replication studies with professional inspectors and more complex industrial requirements are needed to assess the generalizability of these findings and to understand how expertise moderates the impact of LLM support.

\textit{Implications for Teaching.}
The findings also have important implications for teaching RE. Introducing LLMs too early in inspection training may limit students’ opportunities to develop foundational inspection skills and critical reading abilities. Educators should therefore be cautious when integrating LLMs into RE curricula and consider first grounding students in manual inspection techniques before exposing them to AI-based assistance.
These implications are consistent with recent calls for human-centered software engineering education in the AI era, which warn that premature or unstructured use of generative AI tools may undermine foundational learning, reflective practice, and skill development if not carefully integrated into curricula~\cite{abrahao2025software}.
Complementary empirical evidence emerges from studies of novice interaction with LLMs in educational software engineering contexts. Rahe and Maalej report that programming students using ChatGPT frequently delegated substantial portions of problem-solving to the LLM, often entering repetitive cycles of generating, copying, and submitting solutions without engaging in deeper reasoning~\cite{Rahe_2025}. Their analysis also shows that increased interaction time did not translate into improved performance. Despite longer time on task and high perceived usefulness, no corresponding learning or effectiveness gains were observed. These findings closely mirror our observations in RI, reinforcing the concern that unstructured LLM support may reduce cognitive engagement and hinder learning in novice users rather than enhancing inspection effectiveness.

At the same time, LLMs can still play a valuable pedagogical role if used deliberately, for example, as tools for discussion, reflection, or comparison after an initial manual inspection. Used in this way, LLMs may help students articulate reasoning, understand alternative interpretations, and reflect on quality issues without replacing essential learning processes.

\section{Threats to Validity}
\label{sec:threats}
We discuss the threats to validity that apply to this 
study following Wohlin et al.~\cite{wohlin2012experimentation}.

\textbf{Construct Validity.}
\label{sec:threats:construct}
Several covariates (e.g., proficiency in using LLMs and proficiency in RI) may suffer from limited construct validity. These self-reported Likert-scale variables are coarse proxies for the underlying constructs of interest.
Given the difficulty of operationalizing such latent concepts, we resorted to common practice of self-reporting on coarse scales.
The absence of significant effects for these covariates on the outcome variables may therefore stem from this limited construct validity, and we refrain from drawing conclusions about their impact on the experimental task.
The construction of the ground truth may also threaten construct validity. Although the annotation process involved independent inspections by multiple authors followed by iterative consensus-oriented reconciliation discussions, no formal inter-rater agreement metrics were computed. We mitigated this threat through collective review and discussion until a shared agreement was reached for each requirement.

\textbf{Internal Validity.}
\label{sec:threats:internal}
The data collection did not include explicit measures of participants’ general \textit{skill} or \textit{motivation}, both of which are likely to affect all outcome variables.
In RQ2 and RQ3, this threatens internal validity, as skill or motivation may confound the relationship between duration and the other outcome variables.
Because highly skilled or motivated students are likely to both perform faster and better, the absence of an explicit variable for skill or motivation induces a spurious association between those variables~\cite{mcelreath2018statistical}. \Cref{fig:result:rq2:duration} supports this interpretation: longer task durations are counterintuitively associated with lower detection accuracy. We therefore assume this association to be non-causal and confounded by skill or motivation. Since such latent constructs are difficult to operationalize and using proxies such as student grades raises ethical concerns, we chose not to collect this data.
%
Allowing participants to continue the inspection task outside the classroom may also have introduced variability in working conditions and reduced experimental control. 
However, this choice was made to prioritize data completeness and 
avoid comparisons between fully and partially completed inspections. In practice,  this occurred only for two participants during period~1 (2.9\% of the total inspection task executions).

\textbf{External Validity.}
\label{sec:threats:external}
The obtained results are subject to several threats to external validity, which limit the scope of the conclusions. Specifically, the experiment does not fully reflect real RI practice in three dimensions: the participants (i.e., experimental subjects), the requirements (i.e., experimental objects), and the RI task itself (i.e., the experimental task).
Participants were bachelor students in Computer Engineering, without extensive practical experience in RI. While ``using students as participants remains a valid simplification of reality needed in laboratory contexts''~\cite{falessi2018empirical}, the observed effects may differ for experienced industrial practitioners with established inspection strategies and domain knowledge.
Similarly, the requirements used for inspection were simpler than industrial requirements to remain comprehensible within the scope of the experiment. In addition, each requirement contained at most one smell. While this simplified annotation and performance analysis, industrial requirements may contain multiple interacting smells, limiting generalizability to more complex settings.
Finally, the RI task was limited to the detection and classification of requirements smells and relied on a predefined set of smell types rather than fully open-ended identification. Although these smell categories were selected from empirically validated taxonomies grounded in industrial practice~\cite{Frattini2022}, this choice may have influenced participants’ inspection behavior and simplified the task compared to real-world RI settings.
These simplifications were necessary to make the experiment feasible, but limit the external validity of the results beyond novice inspectors in controlled educational contexts.

\textbf{Conclusion Validity.}
\label{sec:threats:conclusion}
The conclusion validity of our results is threatened by two aspects of the data analysis.
Firstly, the sample size of 34 participants is relatively limited.
Secondly, the analysis involves subjective choices, including the selection of a probability distribution family with maximum entropy, linear model parameters, and prior probability distributions.
However, Bayesian data analysis mitigates these threats in two ways. First, it naturally preserves uncertainty in the results, which is reflected in the credibility intervals~\cite{furia2019bayesian}. Second, it makes all analysis steps explicit, including subjective decisions, allowing other researchers to review and revise them if necessary.


\section{Conclusions}\label{sect:conclusion}
This paper presented a controlled crossover-design experiment investigating the effects of LLM support on human-performed RI. Focusing on requirements smell identification, nocuous/innocuous classification, and inspection time, the study provides evidence that LLM support does not necessarily improve inspection outcomes for novice inspectors. In particular, the results show that LLM support negatively affects smell detection accuracy, has no significant effect on smell classification or task duration, and is associated with a reduced learning effect when used from the outset. These findings suggest that early reliance on LLMs may hinder, rather than support, the development of systematic inspection skills.

Future work should investigate alternative ways of integrating LLMs into RI workflows, such as constrained, staged, or post-hoc usage, to mitigate the negative effects observed in this study. Replications with professional inspectors and more complex industrial requirements are needed to assess the generalizability of the findings and to understand how expertise moderates the impact of LLM support. Further research should also explore how different interaction designs, prompting strategies, and explanation mechanisms influence both inspection effectiveness and learning, with the goal of identifying configurations in which LLMs can support human inspection skills.

\section{Data Availability} The replication package is publicly available at~\cite{anonymous_2026_18360108}.



\bibliography{paper/bibliography}

\end{document}